\pdfoutput=1
\documentclass[11pt]{article}
\usepackage[margin=1in]{geometry}
\usepackage[T1]{fontenc}\usepackage{lmodern}\usepackage{microtype}
\usepackage{graphicx,booktabs,amsmath,siunitx}
\usepackage[authoryear,round]{natbib}
\usepackage{xurl}
\usepackage[hidelinks]{hyperref}
\graphicspath{{./}}
\InputIfFileExists{cifras.tex}{}{\newif\ifsintetico
  \newcommand{\cifra}[1]{\textbf{[##1]}}\newcommand{\cifraopc}[1]{--}\newcommand{\cifraabs}[1]{\textbf{[##1]}}\newcommand{\pval}[1]{\textbf{[##1]}}}
\ifsintetico
  \usepackage{draftwatermark}
  \SetWatermarkText{SYNTHETIC --- NOT RESULTS}\SetWatermarkScale{0.32}\SetWatermarkLightness{0.82}\SetWatermarkAngle{45}
\fi
\title{The Delegation Divide: An Open Index of AI-Agent Delegability across Occupations and Countries}
\author{Cristian Espinal Maya\thanks{Universidad Nacional de Colombia, Facultad de Minas. ORCID 0009-0000-1009-8388. Corresponding author.}
  \and Santiago Jim\'enez Londo\~no\thanks{Universidad EAFIT. ORCID 0009-0007-9862-7133.}}
\date{\today}
\begin{document}\maketitle
\begin{abstract}
\cifra{titular.global.l2}\% of task time in world employment is at least partly delegable to AI agents in 2026, and only \cifra{titular.global.l3}\% can be handed over in full.
Existing measures score \emph{exposure} to generative AI, mostly for the United States.
We build the Agent Delegability Index (ADI): two language models rate all 18,796 O*NET 30.2 task statements on a frozen, preregistered rubric of delegation levels and barriers; tasks are mapped to 433 ISCO-08 unit groups and weighted by ILOSTAT employment in \cifra{cobertura.paises} countries.
Four findings.
First, delegable employment rises with income, from \cifra{titular.ingreso.bajo.l2}\% in low-income to \cifra{titular.ingreso.alto.l2}\% in high-income countries, a slope of \cifra{h1.pendiente} percentage points per log point of GDP per capita.
This \emph{delegation divide} is compositional: the physical barrier dominates the jobs poorer countries employ most.
Second, at the occupation level ADI tracks exposure closely ($r$ = \cifra{h3.r.eloundou} with Eloundou et al.\ and \cifra{h3.r.felten} with Felten et al.), above the preregistered bound of 0.80.
ADI adds delegation levels and barriers on top of exposure, for the world, not a new ranking of occupations.
Third, full delegation is rare and peaks for tasks that take a human about \cifra{h4.pico.minutos} minutes.
Fourth, the preregistered adoption gap is not supported: across \cifra{h2.n} countries, delegability and observed AI usage correlate at $r$ = \cifra{h2.r.adi_uso}, and the gap between them is not wider in middle-income countries (\pval{h2.brecha.medio.p}).
Model agreement is $\kappa$ = \cifra{agree.kappa.sonnet_haiku}; a blind human rating of \cifra{n.humano} tasks agrees with the model consensus at $\kappa$ = \cifra{agree.kappa.humano.modelo}, and an agentic panel with web search at $\kappa$ = \cifra{agree.kappa.panel.modelo}.
We release the index, the atlas, the prompts and an MCP server at \url{https://github.com/Cespial/delegation-divide}.
\end{abstract}
\section{Introduction}
\label{sec:intro}

\cifra{titular.global.l2}\% of task time in world employment is at least partly delegable to an AI agent in 2026, and only \cifra{titular.global.l3}\% can be handed over in full.
These two numbers are the output of the Agent Delegability Index (ADI), an open index that rates every task statement of the United States Occupational Information Network (O*NET) on how much of it a software agent can execute, maps the ratings to the International Standard Classification of Occupations (ISCO-08), and weights them by employment in \cifra{cobertura.paises} countries.
This paper describes how the index is built, compares it with existing measures, and reports its first finding: a \emph{delegation divide} between rich and poor countries.

Two gaps motivate the index.
The first is conceptual.
The measures that dominate the literature on artificial intelligence and work score \emph{exposure}: whether a large language model can reduce the time a worker needs for a task by at least half \citep{eloundou2024gpts}, or how closely the abilities an occupation requires track progress in AI applications \citep{felten2021occupational, webb2020impact}.
Exposure was the right construct for a chatbot.
It is not the right construct for an agent.
Since 2025, publicly available systems run code, browse the web, operate a desktop, read and write files and work for hours on a multi-step goal \citep{metr2025measuring}, and evaluations have moved from single answers to agent tasks and to deliverables drawn from real occupations \citep{kapoor2024agents, patwardhan2025gdpval}.
The policy question has moved from ``can the model help with this task?'' to ``can the worker hand this task over?''
That is delegation, and the two differ in kind: a model can halve the time a radiologist spends on a report without being able to sign it, and an agent can file a tax return that no model could have sped up much, because the time was never in the drafting.
Observed usage indices measure a third thing again, adoption, which depends on prices, regulation and habit as much as on capability \citep{anthropic2025economic, lee2026delegates, henseke2026exposure}.

The second gap is geographic.
Almost every task-level measure is built on O*NET and its Standard Occupational Classification (SOC), which only the United States uses for its statistics.
The rest of the world reports employment in ISCO-08 \citep{ilo2012isco}.
Studies that reach other countries do so through coarse bridges, usually at the one-digit level, or through a single national crosswalk \citep{gmyrek2023generative, pizzinelli2023labor, cazzaniga2024gen, albanesi2023new, chen2023large, murugan2026jagged, carbonero2023impact}.
The International Labour Organization (ILO) index of generative-AI exposure is the exception that works natively in ISCO-08 \citep{gmyrek2025generative}, but it measures exposure, not delegation.
Nobody has measured delegability to agents in the classification that most countries use.

ADI addresses both gaps.
The conceptual gap turns out to be narrower in the data than in principle: across occupations, the work that is exposed is mostly the work that is delegable (Section~\ref{sec:validacion}).
The difference lies in how far delegation goes, which barrier stops it, and in which countries the work is done.
The design is simple to state.
Each of the 18,796 task statements in O*NET 30.2 \citep{onet2026database} is rated by frontier language models with a fixed rubric at one of four levels: no meaningful role, augmentation, partial delegation or full delegation (Section~\ref{sec:delegacion}).
The rubric is frozen and preregistered; its hash is in Appendix~\ref{app:rubrica}.
The two models agree with each other at quadratic-weighted $\kappa$ = \cifra{agree.kappa.sonnet_haiku}.
A blind human rating of 300 tasks drawn from a stratified validation sample of 1,500 agrees with the model consensus at $\alpha$ = \cifra{agree.alpha.humano.modelo}, against a preregistered acceptance threshold of 0.60, and an agentic panel with web search agrees with it at $\kappa$ = \cifra{agree.kappa.panel.modelo} over the validation sample (Section~\ref{sec:validacion}).
Tasks are weighted by importance and frequency within their O*NET occupation and carried to 433 of the 436 ISCO-08 unit groups through a documented chain of crosswalks (Appendix~\ref{app:crosswalk}).
Occupation scores are then weighted by employment from the ILO statistics database (ILOSTAT) to give a delegability share for each of \cifra{cobertura.paises} countries with occupational employment data from 2019 onward.

We make five contributions, each meant to stand on its own.
\begin{enumerate}
  \item \textbf{The Agent Delegability Index (ADI).} An open, versioned index of AI-agent delegability for 433 ISCO-08 unit groups and \cifra{cobertura.paises} countries, with three variants per occupation (share of task weight at partial delegation or above, share at full delegation, and expected share of working time delegable), bootstrap confidence intervals, and a permutation placebo. The world headline is \cifra{titular.global.l2}\% (partial or full delegation) and \cifra{titular.global.l3}\% (full delegation).
  \item \textbf{A three-layer definition of delegation.} One citable sentence each for augmentation, partial delegation and full delegation, written for agents with tool use rather than for chat models, with four named barriers (physical, accountability, trust, access) recorded separately from the level. Section~\ref{sec:delegacion} gives the definitions and Table~\ref{tab:comparativa} places them against eight existing measures.
  \item \textbf{A task-level crosswalk and country atlas in ISCO-08.} A reproducible bridge from O*NET-SOC 2019 tasks to ISCO-08 unit groups that keeps \cifra{diseno.puente.conjuntos} distinct task sets (an earlier matrix in the first author's prior work collapsed the same occupations into nine), and an atlas of delegable employment for \cifra{cobertura.paises} countries at the finest ISCO level each country reports, with gender, age and informality breakdowns where household microdata allow.
  \item \textbf{Validation and convergence.} Agreement between models and across repeated calls, a blind human rating and an agentic validation panel, and preregistered convergence tests against the exposure indices of \citet{eloundou2024gpts} and \citet{felten2021occupational}. The correlations are $r$ = \cifra{h3.r.eloundou} and \cifra{h3.r.felten}, above the preregistered bound of 0.80: at the occupation level, the delegability that models rate largely coincides with exposure. ADI's own contribution lies in the level scale, the barriers and the global reach, and it correlates less with observed usage ($r$ = \cifra{h3.r.aei}) than with exposure.
  \item \textbf{Open tools for people and for agents.} The full prompts, the raw model outputs, a Python package (\texttt{pip install adi-index}), static JSON per occupation, a \texttt{llms.txt} file and a Model Context Protocol (MCP) server with three tools (\texttt{lookup\_occupation}, \texttt{country\_profile}, \texttt{cite}), so that an AI agent that needs a delegability score can fetch it and cite it (Appendix~\ref{app:citar}).
\end{enumerate}

The first finding we name is the \emph{delegation divide}.
Delegable employment is not spread evenly across the world.
It rises with income: the share of employment in delegable tasks is \cifra{titular.ingreso.alto.l2}\% in high-income countries and \cifra{titular.ingreso.medio.l2}\% in middle-income countries, and a country-level regression of delegability on log GDP per capita at purchasing power parity has a slope of \cifra{h1.pendiente} percentage points per log point (95\% bootstrap interval \cifra{h1.pendiente.ic}).
The reason is composition.
Poorer countries concentrate employment in agriculture, crafts and elementary occupations, where the binding barrier is physical, and agents have no body.
We also preregistered a second edge, an adoption gap: that middle-income countries would use AI less than their delegable task base predicts.
The data do not support it.
Across \cifra{h2.n} countries with observed usage, delegability and usage correlate at $r$ = \cifra{h2.r.adi_uso}, and the standardized gap between them is not wider in middle-income than in high-income countries (\pval{h2.brecha.medio.p}).
Usage so far follows the delegable task base rather than lagging behind it.

The paper proceeds as follows.
Section~\ref{sec:delegacion} defines the three layers and relates delegation to exposure.
Section~\ref{sec:datos} describes the data and Section~\ref{sec:metodo} the method, validation and inference.
Section~\ref{sec:indice} presents the occupational index, Section~\ref{sec:atlas} the country atlas and the divide, and Section~\ref{sec:validacion} the validation, the convergence with exposure and the comparison with observed adoption.
Section~\ref{sec:limites} discusses limitations and ethics, and Section~\ref{sec:conclusion} concludes.
Appendices give the rubric, the crosswalk, and instructions for citing and using the index.

\section{From exposure to delegation}
\label{sec:delegacion}

\subsection{Three layers}
\label{sec:capas}

ADI rates a task, not an occupation, and rates it at one of four levels.
The choice follows the task-based view of technological change, in which technology substitutes for some tasks within a job and complements others \citep{autor2003skill, acemoglu2011skills, autor2015why}; occupation-level scores of automation risk shrink substantially once the variation of tasks within occupations is taken into account \citep{arntz2016risk, nedelkoska2018automation}.
The reference technology is fixed: publicly available AI agents as of October 2026, that is, frontier language models with tool use, code execution, web browsing, computer use and file handling, with no body and no robotics.
The rater judges what is technically feasible with that technology, not what employers do and not what the law allows.
The four levels, in the wording of the frozen rubric (Appendix~\ref{app:rubrica}), are:

\begin{description}
  \item[L0, no meaningful role.] The agent cannot perform or materially speed up any substantial part of the task.
  \item[L1, augmentation.] A human performs the task while the AI assists step by step with drafts, searches or suggestions; the human stays in the loop at each step and produces the output.
  \item[L2, partial delegation.] An agent can execute substantial sub-steps of the task end to end from a goal specification, while a human still has to perform at least one essential step or review and correct the work before use.
  \item[L3, full delegation.] An agent can complete the whole task end to end, to the quality a competent worker would accept, given only the goal and access, so that the human's role is limited to assigning the task and accepting the result.
\end{description}

\begin{figure}[t]
  \centering
  \includegraphics[width=0.9\linewidth]{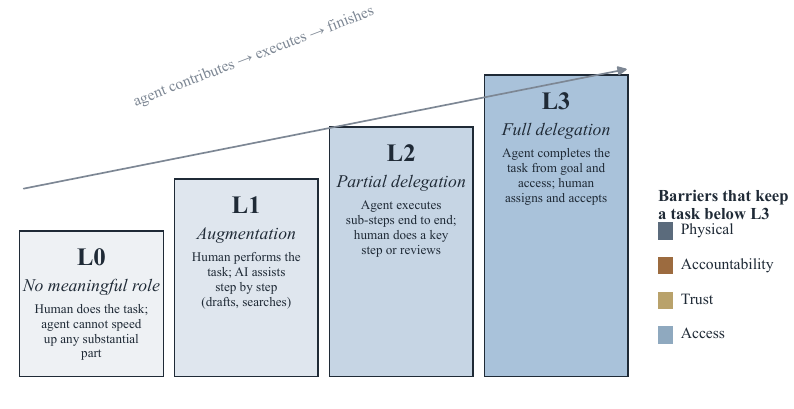}
  \caption{The delegation ladder. Four levels of the ADI rubric, from no meaningful role (L0) to full delegation (L3), with the human and the agent roles at each level and the four barriers that keep a task below L3. Figure generated by \texttt{analisis/figuras.py}.}
  \label{fig:escalera}
\end{figure}

The levels form a ladder (Figure~\ref{fig:escalera}).
Each step adds one thing: at L1 the machine contributes, at L2 it executes, at L3 it finishes.
What moves a task up the ladder is not model quality alone but the shape of the task: whether it can be specified as a goal, whether its inputs and outputs are digital, and whether someone other than the worker is allowed to accept the result.
The rubric therefore records four barriers alongside the level: \emph{physical} (the task needs presence or manipulation), \emph{accountability} (law, licence or liability requires a human to perform or sign), \emph{trust} (the task depends on in-person trust, care or persuasion) and \emph{access} (the task depends on systems, data or places an agent would normally not reach).
It also records the share of the task's working time an agent could take over and the minutes a competent human needs for one instance, which Section~\ref{sec:metodo} uses to test whether full delegation peaks at intermediate task lengths.

\subsection{Why delegation is not exposure}
\label{sec:exposicion}

The exposure indices that most of the literature uses answer a different question.
\citet{eloundou2024gpts} label a task as exposed if access to a large language model, alone or with software built on it, would reduce the time needed to complete the task with equal quality by at least 50\%.
\citet{felten2021occupational} link ten AI application areas to the 52 O*NET abilities and score an occupation by the abilities it needs; the result, the AI Occupational Exposure index (AIOE), does not refer to tasks at all.
\citet{webb2020impact} measures overlap between patent texts and task descriptions, and \citet{brynjolfsson2018what} score each task's suitability for machine learning.
All of them ask how much AI \emph{touches} the work.
ADI asks whether an agent can \emph{do} the work.

The two constructs can separate, in principle, in both directions.
A task can be highly exposed and not delegable: writing a radiology report is faster with a model, but a licensed radiologist must read the images and sign, so the task stays at L1 or L2 with an accountability barrier.
A task can be delegable and barely exposed: reconciling a bank statement against an accounting ledger saved little time when the model could only talk, but an agent with access to both systems completes it end to end.
The experimental evidence on generative AI at work is evidence about this first case: customer-support agents, professional writers and consultants who keep the task and do it faster or better with a model \citep{brynjolfsson2023generative, noy2023experimental, dellacqua2023navigating}.
Exposure measures the intensity of assistance on a task whose owner does not change.
Delegation measures a change of owner.
That the constructs can come apart on a given task does not mean that they do across the economy; Section~\ref{sec:validacion} measures how far they do in practice and finds that, at the level of occupations, they separate little.

Delegation is also not adoption.
The Anthropic Economic Index \citep{anthropic2025economic, handa2025economic} observes which O*NET tasks people actually bring to a model and, in later releases, which they hand to a coding agent; \citet{tomlinson2025working} do the same with conversations with Microsoft Copilot.
The Agentic Adoption Index \citep{lee2026delegates} scores a task by how closely it matches the agent configurations practitioners publish on GitHub, a measure its authors call delegated exposure.
Both record what has happened.
ADI records what the technology permits, so it is an upper bound on delegation that adoption data can be compared against.
The gap between the two is itself an object of study (Section~\ref{sec:validacion}), and it is where the delegation divide appears.

\subsection{Comparison with existing measures}

Table~\ref{tab:comparativa} places ADI next to the measures it builds on.
The columns record what each index measures (construct), the unit it rates, who or what rates it, the occupational classification, the countries it covers, whether the data are open, and whether the construct refers to AI agents rather than to language models or to earlier automation.
ADI is the only task-level index that is natively in ISCO-08, rated for agents, checked against a blind human rating and an agentic panel, and released with tools for agents to query it.

\begin{table}[t]
  \centering
  \footnotesize
  \setlength{\tabcolsep}{3pt}
  \caption{Measures of AI and work compared with ADI. ? = the source does not state it.}
  \label{tab:comparativa}
  \begin{tabular}{@{}p{2.6cm}p{2.5cm}p{1.8cm}p{2.2cm}p{1.9cm}p{1.3cm}cc@{}}
    \toprule
    Measure & Construct & Unit & Rater & Classification & Countries & Open & Agents \\
    \midrule
    \citet{frey2017future} & Probability of computerisation & Occupation (702) & 70 hand-labelled occupations, Gaussian process classifier & SOC 2010 & US & Y & N \\
    \citet{webb2020impact} & Exposure to AI via patent--task text overlap & Task & Text-similarity algorithm & O*NET-SOC & US & Y & N \\
    \citet{felten2021occupational} (AIOE) & Occupational exposure via abilities & Ability (52) & Expert links, AI progress scores & O*NET-SOC & US & Y & N \\
    \citet{eloundou2024gpts} & $\geq$50\% time reduction with LLM ($\alpha$, $\beta$, $\zeta$) & Task & Human annotators and GPT-4 & O*NET-SOC & US & Y & N \\
    \citet{eisfeldt2023generative} & Generative-AI exposure of occupations and firms & Task & ChatGPT & O*NET-SOC & US & ? & N \\
    \citet{gmyrek2025generative} (ILO) & Generative-AI exposure, refined & Task & LLM scores checked by survey and expert review & ISCO-08 & Global via ILO employment & Y & N \\
    \citet{anthropic2025economic} (AEI) & Observed usage of Claude by task & Conversation mapped to task & Automated classifier & O*NET-SOC & US; country series in later releases & Y & Partial \\
    \citet{lee2026delegates} (AAI) & Delegated exposure: match to published agent skills & Task & Embedding similarity to $\sim$888,000 agent skill files & O*NET-SOC & US (GitHub corpus is global) & Y & Y \\
    ADI (this paper) & Delegability to an agent, four levels and four barriers & Task (18,796) & 2 LLMs + blind human rating (300) + agentic panel & ISCO-08 (433 unit groups) & \cifra{diseno.atlas.paises} & Y & Y \\
    \bottomrule
  \end{tabular}
\end{table}

\section{Data}
\label{sec:datos}

\subsection{Tasks: O*NET 30.2}

The unit of rating is the task statement of the Occupational Information Network (O*NET), release 30.2 \citep{onet2026database}.
The release contains \cifra{diseno.onet.tareas} task statements for \cifra{diseno.onet.ocupaciones} O*NET-SOC 2019 occupations, each with its occupation title, and two ratings per task from incumbent surveys: importance (scale 1--5) and frequency (seven categories from yearly or less to hourly or more).
We use the mean importance and the mean frequency category of each task as its weight within its occupation, following the task-weighting convention of \citet{acemoglu2011skills}.
Missing importance or frequency values are imputed with the median, as preregistered.
Each task is rated once; a task that appears under several occupations keeps one rating and several weights.

\subsection{Bridge to ISCO-08}

Most countries report employment in the International Standard Classification of Occupations, ISCO-08 \citep{ilo2012isco}, which has 436 unit groups at four digits.
O*NET uses the United States Standard Occupational Classification (SOC).
We bridge the two in three documented steps: O*NET-SOC 2019 to O*NET-SOC 2010 with the official O*NET Center crosswalk; O*NET-SOC 2010 to six-digit SOC 2010 by truncation; and SOC 2010 to ISCO-08 with the crosswalk of \citet{hardy2018educational}, published by the Institute for Structural Research (IBS).
A SOC code that maps to $k$ ISCO groups splits its weight $1/k$.
The result covers \cifra{diseno.puente.isco} of the 436 unit groups with \cifra{diseno.puente.conjuntos} distinct task sets; the three military groups (0110, 0210, 0310) have no O*NET tasks and are excluded, and seven residual groups without tasks of their own inherit the tasks of their three-digit parent and are flagged as imputed.
Four O*NET-SOC 2019 occupations with no 2010 ancestor are assigned by hand.
Appendix~\ref{app:crosswalk} lists every step, the two SOC codes rescued because the IBS file omits them, the seven imputed groups and the four manual assignments; the code is \texttt{datos/construir\_crosswalk.py}.

The bridge replaces the occupation--task matrix used in the first author's earlier work on augmented human capital \citep{espinal2026augmented, espinal2026measuring}, which mapped SOC to ISCO at one digit and so collapsed \cifra{diseno.previo.ocupaciones} Colombian occupations into \cifra{diseno.previo.conjuntos} task sets.
Section~\ref{sec:limites} returns to the consequences.

\subsection{Secondary task bank: ESCO}

O*NET describes work in the United States.
As a check on that origin, we also score the essential skills and competences that the European Skills, Competences, Qualifications and Occupations classification (ESCO, version 1.2.1) attaches to each of its occupations \citep{esco2026}.
ESCO maps every occupation to an ISCO-08 unit group by construction, so it needs no crosswalk, and it is available in 28 languages.
Its units are skills, not tasks, so we treat it as a robustness bank rather than as the primary source.
We keep only items of the skill and competence type and drop knowledge items, because the rubric rates activities that can be handed over and a body of knowledge is not one.
The bank has \cifraopc{robustez.esco.n.skills} distinct essential skills and competences in \cifraopc{robustez.esco.n.isco} unit groups; each is rated once, by Sonnet alone, with the same rubric and the occupation for which it is most often essential, and enters each unit group with equal weight.
The ESCO results are reported in Section~\ref{sec:atlas} as a sensitivity check and matter most for agriculture, where O*NET has few tasks (Section~\ref{sec:limites}).

\subsection{Employment by occupation: ILOSTAT}

Country weights come from the ILO statistics database, ILOSTAT, series on employment by sex and occupation \citep{ilostat2026}, downloaded through its SDMX interface.
At one digit, \cifra{diseno.ilostat.paises.1d} countries report employment by ISCO-08 major group and \cifra{diseno.ilostat.paises.1d.desde2019} have a reference year of 2019 or later; one of them (Bhutan) reports only armed-forces and unclassified employment in its latest year, so \cifra{diseno.atlas.paises} countries form the atlas.
At two digits, \cifra{diseno.ilostat.paises.2d} countries report sub-major groups and \cifra{diseno.ilostat.paises.2d.desde2019} have a year of 2019 or later.
We use the finest level each country reports, two digits where the sub-major groups are complete (at least thirty groups whose total is within five per cent of the one-digit total) and one digit otherwise, with the pooling rule of Section~\ref{sec:metodo}: \cifra{diseno.atlas.paises.2d} countries enter at two digits and \cifra{diseno.atlas.paises.1d} at one.
Countries whose latest occupational data predate 2019 are excluded, as preregistered.
Table~\ref{tab:cobertura}, generated by \texttt{analisis/tablas.py}, lists coverage by income group.
The same database supplies the informality rate and mean monthly earnings by occupation where they exist, used in the atlas.

\subsection{Income and adoption}

Income groups follow the World Bank classification for fiscal year 2026, and GDP per capita is at purchasing power parity in constant 2021 international dollars \citep{worldbank2026income}.
Observed adoption comes from the Anthropic Economic Index (AEI), which reports usage of Claude by country per working-age adult in its latest release \citep{anthropic2025economic}.
For convergence tests we use the occupation-level scores of \citet{eloundou2024gpts}, \citet{felten2021occupational}, \citet{webb2020impact}, \citet{eisfeldt2023generative} and \citet{gmyrek2025generative}, each carried to ISCO-08 with the same bridge where it is not already in ISCO-08.

\subsection{Household microdata}

For breakdowns by sex, age and informality we use four household surveys that code occupation at four digits in national adaptations of ISCO-08: the Gran Encuesta Integrada de Hogares of Colombia (GEIH) \citep{dane2026geih}, the Pesquisa Nacional por Amostra de Domic\'ilios Cont\'inua of Brazil (PNADC) \citep{ibge2026pnadc}, the Encuesta Nacional de Ocupaci\'on y Empleo of Mexico (ENOE) \citep{inegi2026enoe} and the Encuesta Nacional de Hogares of Peru (ENAHO) \citep{inei2026enaho}.
For Colombia we also use the 2023--2026 monthly series to relate delegability to wage dynamics (Section~\ref{sec:validacion}).
Every input file is listed with its SHA-256 hash and download date in \texttt{datos/MANIFIESTO.jsonl}.

\begin{table}[t]
\centering\small
\caption{Coverage of the country atlas: countries with ILOSTAT employment by ISCO-08 occupation for a year from 2019 onward, by World Bank income group (FY2026). Resolution is the finest ISCO-08 level the country reports.}
\label{tab:cobertura}
\begin{tabular}{lrrrrrr}
\toprule
Income group & Countries & 2-digit & 1-digit & Median year & Employment (M) & With GDP pc \\
\midrule
High & 53 & 46 & 7 & 2024 & 634 & 53 \\
Upper-middle & 43 & 37 & 6 & 2025 & 615 & 43 \\
Lower-middle & 33 & 28 & 5 & 2024 & 916 & 33 \\
Low & 17 & 15 & 2 & 2022 & 170 & 17 \\
\midrule
Total & 146 & 126 & 20 & 2024 & 2,336 & 146 \\
\bottomrule
\end{tabular}
\end{table}

\section{Method}
\label{sec:metodo}

\begin{figure}[t]
  \centering
  \includegraphics[width=\linewidth]{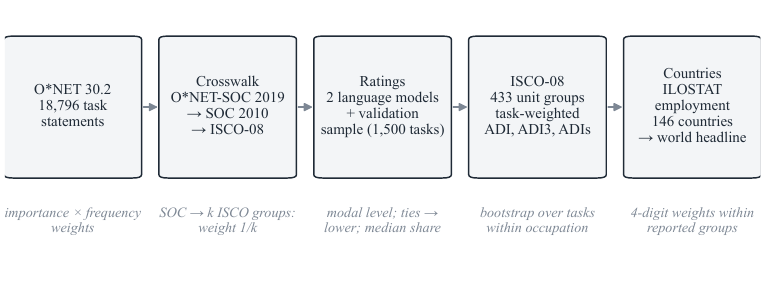}
  \caption{From task statements to a country atlas. Inputs on the left, validation steps below the main flow. Figure generated by \texttt{figuras/fig2\_flujo.py}.}
  \label{fig:flujo}
\end{figure}

Figure~\ref{fig:flujo} summarises the pipeline: tasks are rated by models, consensus is taken across models, consensus is weighted into occupations, occupations are weighted into countries, and every step is validated or bootstrapped.
The design was preregistered on the Open Science Framework on 21 October 2026, before the first call of the main pass, with an embargo until 31 January 2027 \citep{espinal2026prereg}; the hypotheses, the rater set, the rubric hash and the sample seeds are in the registration (\url{https://osf.io/9zntf}), and the exact rater configuration is in \texttt{docs/MODELOS.md}.

\subsection{Rubric}

Each task is rated with the rubric reproduced in Appendix~\ref{app:rubrica}.
The prompt contains the reference technology, the four level definitions, the occupation title and the task text, and asks for a single JSON object with eight fields: level (0--3), share of the task's working time an agent could take over (0--100), minutes a competent human needs for one instance, four barrier flags (physical, accountability, trust, access) and a confidence label.
The rubric contains no example tasks, so no task in the data can anchor the rating of another.
The file was frozen at registration; its SHA-256 hash is \texttt{bcd076fb\ldots0ea0e46} (full value in Appendix~\ref{app:rubrica}), and the scoring script refuses to run the main pass unless the file on disk matches it.
A pilot of at most 300 tasks (seed 20261020) checked JSON compliance, cost and the call format before registration; pilot ratings are discarded (Section~\ref{sec:evaluadores}).

\subsection{Raters: language models}
\label{sec:evaluadores}

Two models rate every task: Claude Sonnet 5.5 is the primary rater and Claude Haiku 5.5 the secondary rater, both by Anthropic (identifiers returned by the interface: \texttt{claude-sonnet-5-5} and \texttt{claude-haiku-5-5}).
The rater set was fixed in the registration and no model is added or dropped after seeing the main pass.
Both are queried through the Claude Code command-line interface (\texttt{claude -p}, version \texttt{2.1.306}, \texttt{-{}-output-format json}) rather than through a raw API, in a neutral configuration: no tools (\texttt{-{}-tools ""}), no Model Context Protocol servers (\texttt{-{}-strict-mcp-config}), no session persistence (\texttt{-{}-no-session-persistence}) and a fixed one-line system prompt that replaces the interface's default, quoted here in full:
\begin{quote}\small\texttt{You are an expert rater of occupational tasks. Follow the user instructions exactly.}\end{quote}
Each call carries one task, with no other task or rating in context.
Every call is logged with the interface version, the model identifier returned by the interface, the prompt hash, the duration and the outcome in \texttt{docs/llm-llamadas.jsonl}, which doubles as the manuscript's declaration of AI use; the configuration is recorded in \texttt{docs/MODELOS.md}.

The interface does not expose a temperature setting, so the ratings are not deterministic by construction.
We therefore measure stability directly: a random subsample of \cifra{n.testretest} tasks is rated a second time by each model under identical prompts, and we report exact agreement and quadratic-weighted kappa between the two passes (Sonnet: \cifra{agree.testretest.sonnet.kappa}; Haiku: \cifra{agree.testretest.haiku.kappa}).
Calls whose output fails validation (malformed JSON, a level outside 0--3, a share outside 0--100) are retried up to five times; tasks that still fail are excluded for that model only, and the counts are reported (\cifra{n.fallos.sonnet} for Sonnet, \cifra{n.fallos.haiku} for Haiku).

The call format was settled in the discarded pilot by two tests, both run before registration.
First, rating tasks in batches of \cifra{diseno.piloto.lote.tareas} per call agreed with one-task calls at quadratic-weighted $\kappa$ = \cifra{diseno.piloto.lote.sonnet.kappa} for Sonnet and \cifra{diseno.piloto.lote.haiku.kappa} for Haiku, below the preregistered floor of 0.80, so batching was rejected and the main pass uses one task per call.
Second, the interface's default wrapper, which tells the model that it is a coding agent with tools, raised the mean level from \cifra{diseno.piloto.neutro.nivel} under the neutral configuration to \cifra{diseno.piloto.envoltorio.nivel} (agreement between the two, $\kappa$ = \cifra{diseno.piloto.envoltorio.kappa}).
The main pass therefore uses the neutral configuration, and a preregistered sensitivity analysis re-rates the 1,500 tasks of the validation sample with the default wrapper (Sections~\ref{sec:validacion_humana} and~\ref{sec:validacion}).

\subsection{Aggregation}

Let $t$ index tasks, $m$ models, $o$ ISCO-08 unit groups and $c$ countries.
The consensus level of a task is the level both models give when they agree and the lower of the two levels when they do not (the preregistered modal rule with ties resolved downward, applied to two raters); the consensus share is the median, here the mean, of the two models' shares.
Task weights $w_{ot}$ are importance times mean frequency, normalised within the O*NET occupation, multiplied by the crosswalk weight ($1/k$ for a SOC code that maps to $k$ ISCO groups) and renormalised so that $\sum_t w_{ot} = 1$ within each unit group.
The primary occupation index is the share of task weight at partial delegation or above:
\begin{equation}
  \mathrm{ADI}_o = \sum_{t} w_{ot}\, \mathbf{1}[\mathrm{level}_t \geq 2],
  \label{eq:adi_o}
\end{equation}
with two preregistered variants, the share at full delegation and the expected share of working time delegable:
\begin{equation}
  \mathrm{ADI3}_o = \sum_{t} w_{ot}\, \mathbf{1}[\mathrm{level}_t = 3], \qquad
  \mathrm{ADIs}_o = \sum_{t} w_{ot}\, \frac{\mathrm{share}_t}{100}.
  \label{eq:adi_var}
\end{equation}
The country index is the employment-weighted mean over the occupational groups $g$ the country reports:
\begin{equation}
  \mathrm{ADI}_c = \frac{\sum_{g} e_{cg}\, \mathrm{ADI}_g}{\sum_{g} e_{cg}},
  \label{eq:adi_c}
\end{equation}
where $e_{cg}$ is ILOSTAT employment and $g$ runs over two-digit sub-major groups if the country reports them for a year from 2019 onward and over one-digit major groups otherwise.
Within a reported group, unit-group indices are averaged with employment weights from microdata of a country in the same World Bank income group, as preregistered. No country publishes employment at the four-digit level in ILOSTAT, so we borrow the within-group structure from two sources mapped through the same bridge as the tasks: United States Occupational Employment and Wage Statistics for May 2024 \citep{bls2025oes} for high-income countries, and Colombia's household survey (GEIH, 2024--2026) \citep{dane2026geih} for middle- and low-income countries. The second is a stated loan for lower-middle and low-income countries, where no four-digit microdata were available to us. Two sensitivity checks are reported: equal weights within each group (preregistered) and GEIH weights for every country.
The world headline is the employment-weighted mean of $\mathrm{ADI}_c$ over the \cifra{diseno.atlas.paises} included countries, and the income-group figures are the same mean within each group.

\subsection{Human and agentic validation}
\label{sec:validacion_humana}

The registration specifies three human raters who each rate the same 1,500 tasks blind \citep{espinal2026prereg}.
Before any result of the main pass was opened, we replaced that design for reasons of time and cost; the change is dated and justified as deviation D2 in \texttt{docs/DESVIACIONES.md}.
The validation now has two parts.

\emph{Blind human rating.}
A subsample of 300 of the 1,500 validation tasks, stratified by ISCO-08 major group with at least 20 tasks per group (seed 20261025; \texttt{validacion/santiago/muestra\_300.csv}), was rated blind by a group led by Santiago Jim\'enez Londo\~no, a coauthor, with the help of student raters.
The 1,500-task sample is the registered one, stratified by major group in proportion to world employment (\texttt{validacion/muestra.py}, seed 20261020).
The raters worked from the frozen rubric and the rater guide, were trained on 20 practice tasks outside the sample, rated in a spreadsheet that contained no model or panel output and no result of the paper, and did not use AI tools while rating.
Which rater scored which task was not recorded, so the sheet is analysed as one human rating.
We report Krippendorff's alpha for ordinal data \citep{krippendorff2004reliability} and quadratic-weighted kappa \citep{cohen1968weighted} between that rating and the model consensus ($\alpha$ = \cifra{agree.alpha.humano.modelo}; $\kappa$ = \cifra{agree.kappa.humano.modelo}; exact agreement \cifra{agree.exacto.humano.modelo.pct}\%; \cifra{n.humano} tasks), and kappa with each model separately (Sonnet \cifra{agree.kappa.humano.sonnet}; Haiku \cifra{agree.kappa.humano.haiku}).
The preregistered acceptance threshold for the primary rater, $\alpha \geq 0.60$, is kept and results are reported whatever the value.
Two features of this design are limitations, not details.
The rating was led by an author of the paper, not by independent raters in the sense of the registration; and because individual contributions were not recorded, no human--human agreement can be computed, so the human figures carry no estimate of how much humans disagree among themselves.
With 300 instead of 1,500 tasks the confidence interval of alpha is also wider than the registered design allowed for.

\emph{An agentic panel, reported as non-human validation.}
Claude Opus 5.5, a model that plays no part in the index, rates the 1,500 validation tasks through the same command-line interface (\texttt{claude -p}) with web search enabled, at most four searches per task; it applies the same rubric and returns up to three pieces of cited evidence with each rating (\texttt{validacion/agentico/panel.py}).
The panel tests whether a rater that can look up how a task is actually done, and what rules govern it, reaches the levels that the raters without tools assign.
We report its agreement with the model consensus, with each model and, on the tasks they share, with the human rating.
The panel is a language model from the same developer as the raters, so it is not human validation and is never reported as such.

The registered design, three human raters on all 1,500 tasks, is left for a second version of the paper.

\emph{Sensitivity to the interface wrapper.}
As preregistered, each model re-rates the 1,500 validation tasks under the interface's default system prompt, which describes the model as a coding agent with tools, instead of the neutral configuration (\texttt{scoring/salidas/envoltorio\_cc/}).
We report, model by model and on the same tasks, the mean level and the share at L2 or above under each configuration, quadratic-weighted kappa between the two, and the agreement of each with the human rating.

\subsection{Convergent validity}

Hypothesis H3 states that occupation-level ADI correlates positively with the exposure scores of \citet{eloundou2024gpts} and the AIOE of \citet{felten2021occupational}, with Pearson $r$ between 0.40 and 0.80 for each: above 0.80 would mean ADI adds little beyond exposure, below 0.40 would question its validity.
Both comparison indices are carried to ISCO-08 with the bridge of Section~\ref{sec:datos}.
We report Pearson and Spearman coefficients (Eloundou: $r$ = \cifra{h3.r.eloundou}, $\rho$ = \cifra{h3.rho.eloundou}; Felten: $r$ = \cifra{h3.r.felten}, $\rho$ = \cifra{h3.rho.felten}), and, as exploratory comparisons, the same statistics against \citet{webb2020impact}, \citet{eisfeldt2023generative}, the ILO index \citep{gmyrek2025generative} and the Agentic Adoption Index \citep{lee2026delegates}.

\subsection{Hypothesis tests}

H1, the delegation divide, is an ordinary least squares regression of $\mathrm{ADI}_c$ on log GDP per capita at purchasing power parity, and a second specification with income-group dummies with high income as reference; the test on the slope is one-sided at $\alpha$ = 0.05.
H2, the adoption gap, standardises $\mathrm{ADI}_c$ and AEI usage per working-age adult to mean zero and unit variance over the countries with both and regresses the difference on income-group dummies; the one-sided test is that the middle-income coefficients are positive.
H4, task length, is a logistic regression of full delegation on log human minutes and its square at the task level, with standard errors clustered by occupation and minutes winsorised at 1 and 4,800; the hypothesis is a positive linear and a negative quadratic coefficient.

\subsection{Inference and placebo}

Uncertainty from the rating step is propagated by a nonparametric bootstrap over tasks: 1,000 draws, tasks resampled within occupation, each draw recomputing Equations~\ref{eq:adi_o}--\ref{eq:adi_c}, the headline and the regression coefficients \citep{efron1993introduction}.
All intervals in the paper are 95\% percentile intervals from this bootstrap unless stated otherwise.
As a placebo, we shuffle task ratings across occupations 1,000 times, holding each occupation's task count and weights fixed, and recompute the occupational index; the observed dispersion of $\mathrm{ADI}_o$ across unit groups is compared with this null (\cifra{placebo.p}).
A rating that carried no occupation-specific information would produce occupational indices indistinguishable from the shuffled ones.
Robustness checks preregistered as exploratory are reported model by model, with ESCO skills instead of O*NET tasks, with uniform task weights, and without the seven imputed groups.

\section{The index}
\label{sec:indice}

Partial delegation is the most common rating of a task, full delegation is rare, and the occupations at the top are clerical and administrative.

\subsection{Tasks}

Of the \cifra{cobertura.tareas.puntuadas} task statements with a valid consensus, \cifra{tareas.nivel0.pct}\% are rated L0, \cifra{tareas.nivel1.pct}\% L1, \cifra{tareas.nivel2.pct}\% L2 and \cifra{tareas.nivel3.pct}\% L3, so that \cifra{tareas.l2.pct}\% of statements reach at least partial delegation before any weighting.
L2 is the modal level.
Agents, as the models judge them, can execute substantial parts of many tasks but can finish very few of them without a human step.
Ties between models, resolved to the lower level as preregistered, affect \cifra{tareas.empates.pct}\% of tasks.
The two models agree with each other at Krippendorff's $\alpha$ = \cifra{agree.alpha.modelos} (ordinal) and at quadratic-weighted $\kappa$ = \cifra{agree.kappa.sonnet_haiku}.

\subsection{Occupations}

Table~\ref{tab:ocupaciones} lists the ten most and the ten least delegable of the \cifra{cobertura.isco} ISCO-08 unit groups covered by the crosswalk (\cifra{cobertura.isco.imputados} of them inherit the tasks of their parent group).
The index spans its whole range.
At the top, \cifra{extremo.isco.n.max} unit groups reach the maximum of \cifra{extremo.isco.top1.l2}\%; ranked within them by the share-based index, the first three are \cifra{extremo.isco.top1.nombre} (ISCO \cifra{extremo.isco.top1.id}); \cifra{extremo.isco.top2.nombre} (\cifra{extremo.isco.top2.id}); and \cifra{extremo.isco.top3.nombre} (\cifra{extremo.isco.top3.id}). The tied groups also include payroll clerks, accounting associate professionals, personnel clerks and database designers and administrators.
At the bottom, \cifra{extremo.isco.n.min} unit groups sit at the minimum of \cifra{extremo.isco.bot1.l2}\%, led, by the share-based index, by \cifra{extremo.isco.bot1.nombre} (\cifra{extremo.isco.bot1.id}); \cifra{extremo.isco.bot2.nombre} (\cifra{extremo.isco.bot2.id}); and \cifra{extremo.isco.bot3.nombre} (\cifra{extremo.isco.bot3.id}); the tied groups also include stonemasons, concrete placers, sewing and embroidery workers and shoemakers.
Full delegation is concentrated in a few clerical groups: coding and proof-reading clerks, typists and word processing operators, contact centre salespersons, receptionists and data entry clerks have the largest shares at L3.

\begin{figure}[t]
  \centering
  \includegraphics[width=0.92\linewidth]{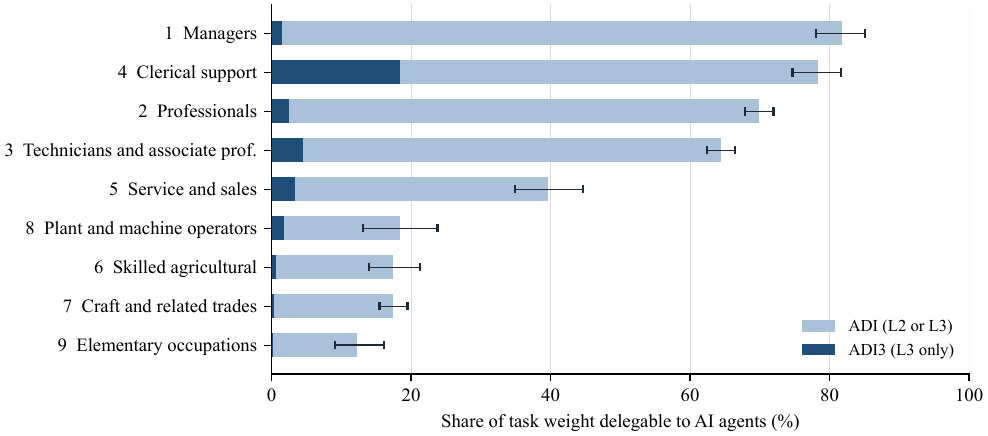}
  \caption{ADI by ISCO-08 major group. Light bars: share of task weight at L2 or L3 (ADI); dark bars: share at L3 (ADI3). Whiskers: 95\% bootstrap interval of ADI (1,000 draws, tasks resampled within occupation). Unit groups are averaged with implied world employment weights. Generated by \texttt{analisis/figuras.py}.}
  \label{fig:isco}
\end{figure}

Figure~\ref{fig:isco} aggregates unit groups to the nine major groups with implied world employment weights.
Managers reach \cifra{gg.1.l2}\%, professionals \cifra{gg.2.l2}\%, technicians and associate professionals \cifra{gg.3.l2}\% and clerical support workers \cifra{gg.4.l2}\%.
Service and sales workers sit in the middle at \cifra{gg.5.l2}\%.
The manual groups sit far below: \cifra{gg.6.l2}\% for skilled agricultural workers, \cifra{gg.7.l2}\% for craft and related trades, \cifra{gg.8.l2}\% for plant and machine operators and \cifra{gg.9.l2}\% for elementary occupations.
The order is the one exposure indices produce, with cognitive and office work on top and manual work at the bottom; Section~\ref{sec:validacion} quantifies how close it is.
Two features are specific to delegation.
Managers rank first at L2 or above but, like professionals, have very little weight at L3: the models leave the decision with a human.
Clerical support workers have by far the largest share at L3 of any major group, because their tasks are digital, specified by a goal and accepted by someone else.

\begin{table}[t]
\centering\footnotesize
\caption{The ten most and ten least delegable ISCO-08 unit groups. ADI: task-weighted share at L2 or L3; ADI3: at L3; ADIs: weighted mean share of task time delegable, with 95\% bootstrap interval (1,000 draws, tasks resampled within occupation).}
\label{tab:ocupaciones}
\begin{tabular}{lp{5.6cm}rrrcr}
\toprule
ISCO & Unit group & ADI & ADI3 & ADIs & 95\% CI (ADIs) & Tasks \\
\midrule
\multicolumn{7}{l}{\emph{Most delegable}}\\
4413 & Coding, proof-reading and related clerks & 100.0 & 67.7 & 79.6 & [73.8, 83.9] & 11 \\
5244 & Contact centre salespersons & 100.0 & 43.1 & 79.0 & [76.1, 81.8] & 12 \\
4419 & Clerical support workers not elsewhere classified & 100.0 & 41.1 & 75.0 & [70.9, 79.0] & 36 \\
4313 & Payroll clerks & 100.0 & 16.1 & 73.9 & [71.3, 76.5] & 21 \\
3313 & Accounting associate professionals & 100.0 & 35.5 & 73.3 & [68.9, 77.8] & 28 \\
4416 & Personnel clerks & 100.0 & 22.4 & 70.4 & [65.8, 75.4] & 19 \\
4322 & Production clerks & 100.0 & 13.4 & 65.6 & [60.4, 70.5] & 17 \\
4323 & Transport clerks & 100.0 & 8.3 & 65.5 & [61.3, 70.4] & 12 \\
2521 & Database designers and administrators & 100.0 & 6.4 & 64.6 & [62.3, 67.0] & 43 \\
4229 & Client information workers not elsewhere classified & 100.0 & 5.0 & 62.6 & [58.9, 66.5] & 17 \\
\midrule
\multicolumn{7}{l}{\emph{Least delegable}}\\
9215 & Forestry labourers & 0.0 & 0.0 & 9.3 & [5.7, 13.4] & 17 \\
8153 & Sewing machine operators & 0.0 & 0.0 & 7.7 & [4.3, 12.2] & 26 \\
8156 & Shoemaking and related machine operators & 0.0 & 0.0 & 7.2 & [2.7, 13.4] & 19 \\
7533 & Sewing, embroidery and related workers & 0.0 & 0.0 & 6.9 & [4.4, 10.3] & 37 \\
7113 & Stonemasons, stone cutters, splitters and carvers & 0.0 & 0.0 & 6.6 & [4.3, 9.2] & 28 \\
8157 & Laundry machine operators & 0.0 & 0.0 & 6.4 & [3.7, 10.8] & 23 \\
7114 & Concrete placers, concrete finishers and related workers & 0.0 & 0.0 & 6.4 & [4.3, 8.8] & 64 \\
9311 & Mining and quarrying labourers & 0.0 & 0.0 & 5.9 & [3.9, 8.0] & 14 \\
9312 & Civil engineering labourers & 0.0 & 0.0 & 5.3 & [4.3, 6.4] & 45 \\
9121 & Hand launderers and pressers & 0.0 & 0.0 & 5.0 & [2.5, 8.9] & 28 \\
\bottomrule
\end{tabular}
\end{table}

\subsection{Barriers}

\begin{figure}[t]
  \centering
  \includegraphics[width=0.92\linewidth]{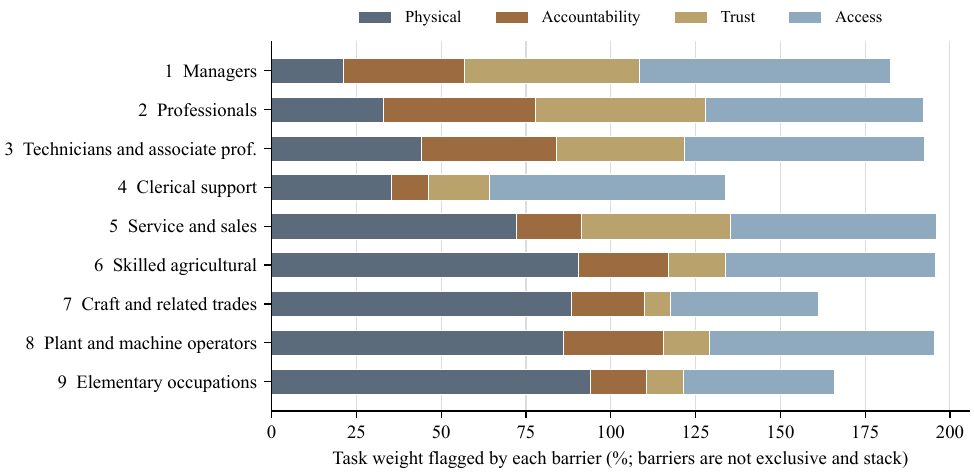}
  \caption{Barriers that keep tasks below full delegation, by ISCO-08 major group: task weight flagged by each barrier (fraction of models that flag it, weighted as in the index). Barriers are not mutually exclusive and stack. Generated by \texttt{analisis/figuras.py}.}
  \label{fig:barreras}
\end{figure}

Figure~\ref{fig:barreras} decomposes the task weight that each barrier flags by major group (exploratory).
Two barriers split the occupational structure.
The physical barrier flags most of the task weight in the five manual and service groups, from \cifra{gg.5.barrera_fisica.pct}\% in service and sales work to \cifra{gg.6.barrera_fisica.pct}\% in agriculture, \cifra{gg.7.barrera_fisica.pct}\% in crafts, \cifra{gg.8.barrera_fisica.pct}\% among operators and \cifra{gg.9.barrera_fisica.pct}\% in elementary occupations, and it is what keeps them low.
In the four office groups the most frequent flag is access, which marks between \cifra{gg.2.barrera_acceso.pct}\% (professionals) and \cifra{gg.1.barrera_acceso.pct}\% (managers) of task weight: the task depends on systems, data or places an agent would not normally reach.
Accountability is highest among professionals (\cifra{gg.2.barrera_responsabilidad.pct}\%), technicians (\cifra{gg.3.barrera_responsabilidad.pct}\%) and managers (\cifra{gg.1.barrera_responsabilidad.pct}\%), and trust among managers (\cifra{gg.1.barrera_confianza.pct}\%) and professionals (\cifra{gg.2.barrera_confianza.pct}\%), where tasks involve signatures, liability, persuasion and care; trust is also frequent in service and sales work (\cifra{gg.5.barrera_confianza.pct}\%).
Clerical work carries the fewest accountability flags of any group (\cifra{gg.4.barrera_responsabilidad.pct}\%) and few trust flags (\cifra{gg.4.barrera_confianza.pct}\%), which is why it is the group that reaches full delegation.
The barriers stack, so a task can carry several; the figure reports each one separately rather than assigning a single cause.

\subsection{Task length (H4)}

H4 is supported: full delegation peaks for tasks that take a human about twelve minutes.
The probability that a task is fully delegable follows an inverted U in the log of the minutes a human needs for it: the logistic coefficients are \cifra{h4.b1} on log minutes (one-sided \pval{h4.p1}) and \cifra{h4.b2} on its square (\pval{h4.p2}), over \cifra{h4.n} tasks clustered in \cifra{h4.n.clusters} O*NET occupations.
The implied peak is at about \cifra{h4.pico.minutos} minutes of human work.
Very short tasks are often moments inside a physical or interpersonal activity, which an agent cannot isolate and finish on its own.
Long tasks bundle several steps, including judgement, coordination or sign-off, and one human step is enough to keep the whole task at L2.
Full delegation lives in between, in self-contained pieces of digital work.
The minutes are themselves a model estimate, recorded in the same call as the level, so the shape of the curve should be read as the models' view of task length rather than as a measured duration.

\subsection{Robustness}

The occupational index is not an artefact of the task-to-occupation assignment: the dispersion of $\mathrm{ADI}_o$ across unit groups (standard deviation \cifra{placebo.sd.obs}) compares with a mean of \cifra{placebo.sd.media} under the permutation null (\pval{placebo.p}).
With uniform task weights the world headline is \cifra{titular.global.l2.pesos_uniformes}\%, and without the imputed unit groups \cifra{titular.global.l2.sin_imputados}\%.

The choice of model matters more than any weighting choice.
Model by model, the world headline is \cifra{titular.modelo.sonnet.l2}\% with Sonnet and \cifra{titular.modelo.haiku.l2}\% with Haiku.
Haiku is the more generous rater: it places many more tasks at L2 or above.
Because the preregistered consensus takes the lower level when the two models disagree, the consensus headline of \cifra{titular.global.l2}\% stays close to Sonnet, the primary rater, and the gap to Haiku is a measure of how much the index depends on which model is asked.
Both models are stable across repeated calls (Section~\ref{sec:validacion}), so the difference is systematic, not noise.

\section{The atlas: a delegation divide}
\label{sec:atlas}

A minority of task time in world employment is delegable to AI agents, almost none of it in full, and the share rises steeply with national income.

\subsection{The headline}

Weighting each country by its employment, \cifra{titular.global.l2}\% of task time in world employment is delegable to AI agents (95\% bootstrap interval \cifra{titular.global.l2.lo}--\cifra{titular.global.l2.hi}), and \cifra{titular.global.l3}\% is fully delegable (\cifra{titular.global.l3.lo}--\cifra{titular.global.l3.hi}).
The share-based variant, the employment-weighted mean share of task time an agent can carry, is \cifra{titular.global.share}\%.
The headline covers \cifra{cobertura.paises} countries and \cifra{cobertura.empleo_millones} million workers (Table~\ref{tab:cobertura}); \cifra{cobertura.paises.2d} countries report employment at two ISCO digits and \cifra{cobertura.paises.1d} at one.
The headline does not depend on how unit groups are weighted inside the groups a country reports.
With equal weights, the preregistered sensitivity check, it is \cifra{titular.global.l2.iguales}\% (\cifra{titular.global.l3.iguales}\% at L3); with Colombian weights for every country it is \cifra{titular.global.l2.geih}\% (\cifra{titular.global.l3.geih}\% at L3); without any employment weighting of unit groups it is \cifra{titular.global.l2.sin_ponderar}\%.

Four countries were excluded after the first real run because their occupational employment data do not add up to their total employment: Suriname, Senegal, Comoros and Hong Kong (deviation D3 in \texttt{docs/DESVIACIONES.md}).
The rule was set after seeing results, so we also computed the headline and the H1 slope with the original \cifra{d3.n.paises} countries: the headline is \cifra{d3.titular.l2.150}\% (against \cifra{d3.titular.l2.con_regla}\%) and the slope \cifra{d3.h1.pendiente.150} (against \cifra{h1.pendiente}). No conclusion depends on the exclusion.

\begin{figure}[t]
  \centering
  \includegraphics[width=\linewidth]{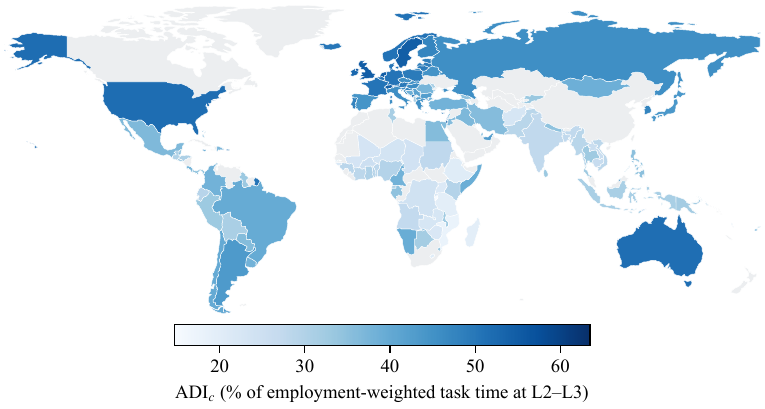}
  \caption{The Agent Delegability Index by country ($\mathrm{ADI}_c$): employment-weighted share of task time at L2 or L3, at the finest ISCO-08 level each country reports to ILOSTAT for a year from 2019 onward. Grey: no occupation data. Boundaries: Natural Earth 1:110m. Generated by \texttt{analisis/figuras.py}.}
  \label{fig:mapa}
\end{figure}

\subsection{Delegability and income (H1)}

\begin{figure}[t]
  \centering
  \includegraphics[width=0.85\linewidth]{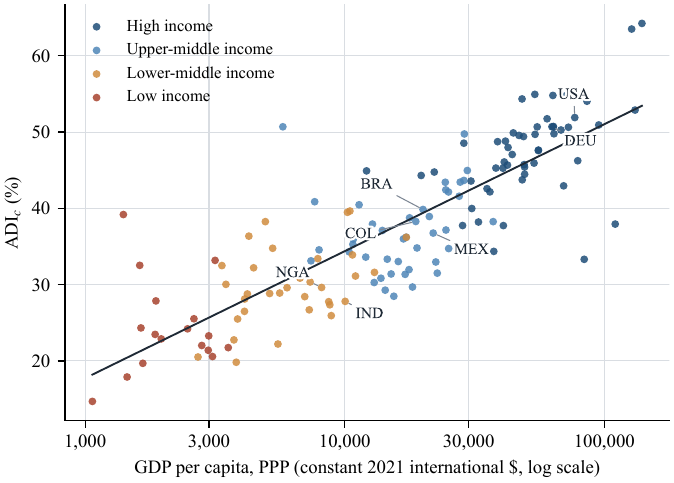}
  \caption{The delegation divide. $\mathrm{ADI}_c$ against GDP per capita at purchasing power parity (log scale), coloured by World Bank income group (FY2026), with the OLS line of the preregistered H1 regression. Generated by \texttt{analisis/figuras.py}.}
  \label{fig:ingreso}
\end{figure}

H1 is supported: delegable employment rises with income.
Figure~\ref{fig:mapa} maps $\mathrm{ADI}_c$ and Figure~\ref{fig:ingreso} plots it against GDP per capita.
The share of employment in delegable tasks is \cifra{titular.ingreso.alto.l2}\% in high-income countries, \cifra{titular.ingreso.medio_alto.l2}\% in upper-middle-income, \cifra{titular.ingreso.medio_bajo.l2}\% in lower-middle-income and \cifra{titular.ingreso.bajo.l2}\% in low-income countries.
The ordering holds at full delegation: \cifra{titular.ingreso.alto.l3}\%, \cifra{titular.ingreso.medio_alto.l3}\%, \cifra{titular.ingreso.medio_bajo.l3}\% and \cifra{titular.ingreso.bajo.l3}\%.
The preregistered regression (Table~\ref{tab:h1}, column 1) has a slope of \cifra{h1.pendiente} percentage points per log point of GDP per capita (one-sided \pval{h1.p}; 95\% bootstrap interval \cifra{h1.pendiente.ic}; $R^2$ = \cifra{h1.r2}; \cifra{h1.n} countries).
With heteroskedasticity-robust errors the standard error is \cifra{h1.se.hc1}, against \cifra{h1.se} under the classical assumption.
With income-group dummies, every group sits below high income in the preregistered direction: low-income countries by \cifraabs{h1.dif.bajo_alto} points (\pval{h1.dif.bajo_alto.p}), lower-middle by \cifraabs{h1.dif.medio_bajo_alto} (\pval{h1.dif.medio_bajo_alto.p}) and upper-middle by \cifraabs{h1.dif.medio_alto_alto} (\pval{h1.dif.medio_alto_alto.p}); the dummies alone give $R^2$ = \cifra{h1.dummies.r2}.
The slope is \cifra{h1.pendiente.iguales} with equal within-group weights, \cifra{h1.pendiente.geih} with Colombian weights for every country and \cifra{h1.pendiente.share} with the share-based index.
A placebo in which tasks are shuffled across occupations reproduces the slope in a fraction \cifra{placebo.h1.p} of draws.
With ESCO essential skills in place of O*NET tasks (exploratory, Sonnet only, same country weights), the world headline is \cifra{robustez.esco.titular.l2}\% and the slope is \cifra{robustez.esco.h1.pendiente} (one-sided \pval{robustez.esco.h1.p}); across unit groups the ESCO index correlates with the O*NET index from Sonnet at $r$ = \cifra{robustez.esco.r.isco}.

The divide is a statement about the composition of employment, and should be read as one.
Every country receives the same occupational index, built from the United States task mix, so differences between countries come from which occupations employ their workers and, at the margin, from the within-group weights of their income group.
Low-income countries employ most of their workers in agriculture, crafts and elementary occupations, where the physical barrier flags most of the task weight (Section~\ref{sec:indice}).
High-income countries employ more managers, professionals and clerks, whose tasks are digital.
The divide therefore does not say that an agent performs the same task worse in a poorer country; it says that the work poorer countries do is work agents cannot reach.
Section~\ref{sec:limites} discusses what the transfer of the United States task mix may hide.

\begin{table}[t]
\centering\small
\caption{Country-level regressions. Columns (1)--(3): $\mathrm{ADI}_c$ in percent (H1); column (3) uses equal weights within reported groups (sensitivity). Column (4): standardized gap $z(\mathrm{ADI}_c)-z(\text{usage}_c)$ with Anthropic Economic Index usage per working-age person (H2). Reference category: high income. Standard errors in parentheses; stars refer to the preregistered one-sided tests (* $p<0.05$, ** $p<0.01$, *** $p<0.001$).}
\label{tab:h1}
\begin{tabular}{lcccc}
\toprule
 & (1) H1 & (2) H1 & (3) H1, equal weights & (4) H2 gap \\
\midrule
log GDP per capita (PPP) & 7.23*** &  & 7.43 &  \\
 & (0.39) &  & (0.39) &  \\
Upper-middle income &  & $-$10.57*** &  & 0.17 \\
 &  & (1.15) &  & (0.12) \\
Lower-middle income &  & $-$17.57*** &  & $-$0.06 \\
 &  & (1.24) &  & (0.13) \\
Low income &  & $-$23.06*** &  & $-$0.81 \\
 &  & (1.56) &  & (0.24) \\
Constant & $-$32.18 & 47.45 & $-$34.69 & 0.01 \\
 & (3.79) & (0.77) & (3.76) & (0.08) \\
\midrule
Countries & 146 & 146 & 146 & 101 \\
$R^2$ & 0.71 & 0.70 & 0.72 & 0.15 \\
\bottomrule
\end{tabular}
\end{table}

\subsection{Countries}

The extremes of the atlas follow the same logic.
The most delegable countries are small, service-intensive, high-income economies: \cifra{extremo.pais.top1.nombre} (\cifra{extremo.pais.top1.l2}\%), \cifra{extremo.pais.top2.nombre} (\cifra{extremo.pais.top2.l2}\%) and \cifra{extremo.pais.top3.nombre} (\cifra{extremo.pais.top3.l2}\%) lead, followed closely by Sweden, the Netherlands and Israel.
The least delegable are agrarian, low- and lower-middle-income economies in Africa: \cifra{extremo.pais.bot1.nombre} (\cifra{extremo.pais.bot1.l2}\%), \cifra{extremo.pais.bot2.nombre} (\cifra{extremo.pais.bot2.l2}\%) and \cifra{extremo.pais.bot3.nombre} (\cifra{extremo.pais.bot3.l2}\%), followed by Tanzania, Lesotho and Ethiopia.
Among large economies, Colombia stands at \cifra{pais.COL.l2}\%, Mexico at \cifra{pais.MEX.l2}\% and Brazil at \cifra{pais.BRA.l2}\%, against \cifra{pais.USA.l2}\% in the United States and \cifra{pais.IND.l2}\% in India.
The three Latin American countries sit above the upper-middle-income average, close to the world headline.
The full country table, with bootstrap intervals and the equal-weight variant, is released as \texttt{salidas/adi\_pais.csv}.

\section{Validation, convergence and adoption}
\label{sec:validacion}

The ratings are stable and the two models largely agree; at the occupation level, ADI coincides with existing exposure indices more than the preregistration anticipated; and the adoption gap we predicted does not appear.

\subsection{Agreement between models and over time}

The two models agree at quadratic-weighted $\kappa$ = \cifra{agree.kappa.sonnet_haiku} and Krippendorff's $\alpha$ = \cifra{agree.alpha.modelos} (ordinal), over \cifra{cobertura.tareas.puntuadas} tasks.
They differ in generosity: Haiku places many more tasks at L2 or above, and its headline is \cifra{titular.modelo.haiku.l2}\% against \cifra{titular.modelo.sonnet.l2}\% for Sonnet.
The preregistered consensus takes the lower level when they disagree, so the headline follows Sonnet, the primary rater; the Haiku figure is an upper reading of the same rubric.
Stability across repeated calls on \cifra{n.testretest} randomly drawn tasks is $\kappa$ = \cifra{agree.testretest.sonnet.kappa} for Sonnet and \cifra{agree.testretest.haiku.kappa} for Haiku, with exact agreement on \cifra{agree.testretest.sonnet.exacto.pct}\% and \cifra{agree.testretest.haiku.exacto.pct}\% of tasks.
The interface exposes no temperature, yet a second call returns the same level for most tasks; the disagreement between models is therefore systematic, not sampling noise.

\subsection{Human and agentic validation}

The preregistered human validation changed before any result was opened (deviation D2; Section~\ref{sec:validacion_humana}).
Instead of three human raters on 1,500 tasks, a subsample of 300 of them was rated blind by a group led by a coauthor with the help of student raters, none of whom saw any model rating or any result of the paper.
Which rater scored which task was not recorded, so the sheet is analysed as a single human rating: it gives no human--human agreement, and a rating led by a coauthor is not independent in the sense of the registration; we report its agreement with the models as a limitation-bound check, not as a validation equivalent to the registered one.
Human--model agreement is $\alpha$ = \cifra{agree.alpha.humano.modelo} (ordinal) and quadratic-weighted $\kappa$ = \cifra{agree.kappa.humano.modelo} over \cifra{n.humano} tasks, against the preregistered acceptance threshold of $\alpha \geq 0.60$ for the primary rater; by model, $\kappa$ is \cifra{agree.kappa.humano.sonnet} for Sonnet and \cifra{agree.kappa.humano.haiku} for Haiku, and the human rating gives the same level as the consensus on \cifra{agree.exacto.humano.modelo.pct}\% of tasks.
The human rating is more generous than the consensus: its mean level is \cifra{humano.nivel.media} against \cifra{modelo.nivel.media.humano}, and it places \cifra{humano.l2.pct}\% of the tasks at L2 or above against \cifra{modelo.l2.pct.humano}\% for the consensus.
Of the \cifra{humano.l3.n} tasks the human rating puts at L3, the consensus puts \cifra{humano.l3.consenso.l2.n} at L2: the disagreement is almost entirely about whether a task an agent can carry out end to end still needs a human to check the result, not about whether it can be delegated at all.
Every figure is reported whatever its value.

The agentic panel, Claude Opus 5.5 with web search, which plays no part in the index, rated all 1,500 validation tasks (\cifra{n.panel} with a model consensus).
On those tasks its agreement with the model consensus is quadratic-weighted $\kappa$ = \cifra{agree.kappa.panel.modelo} and $\alpha$ = \cifra{agree.alpha.panel.modelo} (ordinal), with exact agreement on \cifra{agree.exacto.panel.modelo.pct}\% of tasks; with each model, $\kappa$ = \cifra{agree.kappa.panel.sonnet} for Sonnet and \cifra{agree.kappa.panel.haiku} for Haiku.
Its mean level is \cifra{panel.nivel.media}, against \cifra{modelo.nivel.media.muestra} for the consensus on the same tasks, and it places \cifra{panel.l2.pct}\% of them at L2 or above, against \cifra{modelo.l2.pct.muestra}\% for the consensus.
On the \cifra{n.panel.humano} tasks it shares with the human rating, its agreement with it is $\kappa$ = \cifra{agree.kappa.panel.humano}.
The panel tests whether a rater with tools and evidence reaches the same levels as raters without them.
It is a model from the same developer as the raters, so its agreement with them may partly reflect a shared view of work; it is not human validation and is not reported as such.

\subsection{Convergence with exposure indices (H3)}

H3 is not supported in the form registered, and the reason matters for what ADI is.
Across ISCO-08 unit groups, ADI correlates with the exposure measure of \citet{eloundou2024gpts} at $r$ = \cifra{h3.r.eloundou} (Spearman $\rho$ = \cifra{h3.rho.eloundou}) and with the AIOE of \citet{felten2021occupational} at $r$ = \cifra{h3.r.felten} ($\rho$ = \cifra{h3.rho.felten}).
The preregistration set a band of 0.40 to 0.80 and stated in advance that a correlation above 0.80 would mean that ADI adds little beyond exposure.
Both correlations are above it.
At the level of occupations, the delegability that language models rate largely coincides with the exposure that earlier studies measured: the occupations an agent could take over are, for the most part, the occupations a chatbot could speed up.
The correlation with \citet{eisfeldt2023generative} is similar, at \cifra{h3.r.eisfeldt}.
A ranking of occupations is therefore not ADI's contribution.

What ADI adds lies elsewhere, and four results locate it.
First, the level scale: exposure indices say whether AI touches a task, while ADI separates partial from full delegation, and full delegation turns out to be rare (\cifra{titular.global.l3}\% of world task time) and concentrated in a few clerical groups (Section~\ref{sec:indice}).
Second, the barriers, which say why a task stops short of L3 and which are not recorded by any exposure index.
Third, the reach: a task-level index carried to \cifra{cobertura.paises} countries in ISCO-08, where exposure indices are mostly confined to the United States.
Fourth, the relation to what is observed and to other kinds of automation.
ADI correlates at only \cifra{h3.r.aei} with observed Claude usage by occupation \citep{handa2025economic} (\cifra{h3.r.aei_tareas} when usage is matched task by task), well below its correlation with exposure, so capability and use are not the same ranking.
It is nearly uncorrelated with the AI-patent exposure of \citet{webb2020impact} ($r$ = \cifra{h3.r.webb_ai}), negatively correlated with Webb's software exposure (\cifra{h3.r.webb_software}) and strongly negatively correlated with his robot exposure (\cifra{h3.r.webb_robot}): the work agents can take over is the opposite of the work robots were expected to take.

\subsection{Observed usage and the adoption gap (H2)}

H2 is not supported.
For the \cifra{h2.n} countries with Anthropic Economic Index usage per working-age person \citep{anthropic2025economic}, delegability and usage correlate at $r$ = \cifra{h2.r.adi_uso}.
The preregistered test regresses the standardized gap between the two on income-group dummies.
The gap in middle-income countries is \cifra{h2.brecha.medio} standard deviations relative to high-income countries, not distinguishable from zero (one-sided \pval{h2.brecha.medio.p}).
Separately, the coefficient is \cifra{h2.b.medio_alto} for upper-middle (\pval{h2.b.medio_alto.p}) and \cifra{h2.b.medio_bajo} for lower-middle income (\pval{h2.b.medio_bajo.p}; Table~\ref{tab:h1}, column 4).
Neither is significant, and the lower-middle coefficient has the wrong sign.
The only group whose gap departs clearly from high income is low income, and it departs in the opposite direction to the hypothesis: the coefficient is \cifra{h2.b.bajo} (two-sided \pval{h2.b.bajo.p}), so usage there is less far below the high-income level than delegability is.
We read this as a null result.
Across countries, observed usage of one AI system follows the delegable task base closely; we find no evidence that middle-income countries lag behind what their occupational structure would allow.
Two cautions apply.
Usage comes from a single developer's product and measures conversations, not delegated work, and the same developer's models produce the ratings, so part of the high correlation may reflect a shared view of which work suits these models (Section~\ref{sec:limites}).

\subsection{Sensitivity to the interface wrapper}

The preregistered wrapper sensitivity re-rates the 1,500 validation tasks under the default Claude Code system prompt, which describes the model as an agent with tools.
In the pilot, that wrapper raised the mean level (Section~\ref{sec:evaluadores}).
For Haiku, over the \cifra{envoltorio.haiku.n} re-rated tasks, the mean level moves from \cifra{envoltorio.haiku.nivel.neutro} under the neutral configuration to \cifra{envoltorio.haiku.nivel.cc} under the default wrapper, the share at L2 or above from \cifra{envoltorio.haiku.l2.neutro.pct}\% to \cifra{envoltorio.haiku.l2.cc.pct}\%, and agreement between the two configurations is $\kappa$ = \cifra{envoltorio.haiku.kappa}.
For Sonnet, the primary rater, the same figures are \cifra{envoltorio.sonnet.nivel.neutro} and \cifra{envoltorio.sonnet.nivel.cc}, \cifra{envoltorio.sonnet.l2.neutro.pct}\% and \cifra{envoltorio.sonnet.l2.cc.pct}\%, with $\kappa$ = \cifra{envoltorio.sonnet.kappa} (\cifra{envoltorio.sonnet.n} tasks): the wrapper moves Sonnet far more than Haiku.
Against the human rating, the neutral and default configurations reach $\kappa$ = \cifra{envoltorio.sonnet.kappa.humano.neutro} and \cifra{envoltorio.sonnet.kappa.humano.cc} for Sonnet, and \cifra{envoltorio.haiku.kappa.humano.neutro} and \cifra{envoltorio.haiku.kappa.humano.cc} for Haiku.
The frame in which a model is addressed therefore enters the index; Section~\ref{sec:limites} discusses what this means for reading it.

\section{Limitations and ethics}
\label{sec:limites}

\subsection{Technical delegability is not deployment}

ADI measures what an agent can do with the technology of October 2026, given the goal and access.
It does not measure what firms do, what regulators allow or what workers accept.
A task at L3 may never be delegated because the agent is too expensive relative to local wages, because the systems it would need are not connected, or because nobody is willing to be accountable for its output.
The index is therefore an upper bound on delegation, and the comparison with observed usage in Section~\ref{sec:validacion} is a comparison of a ceiling with a floor.
Adoption of generative AI has been fast among workers \citep{bick2024rapid}, but its measured effects on earnings and hours have so far been small \citep{humlum2025large}, with early employment losses concentrated among young workers in exposed occupations \citep{brynjolfsson2025canaries}.
Readers who need a forecast of employment effects should combine ADI with adoption, prices and institutions, as \citet{acemoglu2024simple} does for exposure; the index alone supports none.
The reference technology is also a moving target.
We fix it by date and version the index (\texttt{v2026.1}) so that later releases can be compared with this one rather than silently replacing it.

\subsection{O*NET describes work in the United States}

The task statements, and the importance and frequency weights, come from surveys of United States workers.
The same occupation elsewhere may contain different tasks, in different proportions, with different tools.
The crosswalk to ISCO-08 carries the United States task mix to every country; it does not adapt it.
The problem is sharpest in agriculture.
O*NET has only \cifra{diseno.agro.tareas} task statements whose main ISCO group lies in major group 6, skilled agricultural, forestry and fishery workers, which employs a large share of workers in low-income countries, and \cifra{diseno.agro.en_muestra} of them fall in the human-validation sample.
Since the delegation divide (H1) depends on that group, we rerun the country regression with the ESCO skill bank in place of O*NET tasks for major group 6 and for all groups; the slope is \cifra{robustez.esco.h1.pendiente} with ESCO for all groups and \cifra{robustez.esco.h1.pendiente.gg6} with ESCO for major group 6 only, against \cifra{h1.pendiente} with O*NET.
The same transfer problem is known from the measurement of teleworkability: a task-based feasibility measure built on O*NET \citep{dingel2020how} overstates what can be done at home in poorer countries, where the same occupations are performed with fewer digital tools \citep{gottlieb2021working}, and the access barrier in ADI plays the role that connectivity plays there \citep{gmyrek2024buffer}.
Informal and subsistence work, which household surveys capture better than labour-force statistics by occupation, is treated in the microdata breakdowns rather than in the headline.

\subsection{Language models as raters}

The raters are two language models from a single developer, and the usage data in Section~\ref{sec:validacion} come from the same developer's Anthropic Economic Index.
Three risks follow.
The first is bias: a model may rate tasks in its own domain, such as software, as more delegable than tasks it knows less well, and may share the blind spots of its training data about work outside rich countries.
The human check that would bound this bias is weaker than the one we registered (deviation D2).
The human rating of 300 tasks was led by a coauthor with the help of student raters: they were blind to the model ratings and to the results, but the lead shares the authors' framing of the question, and because which rater scored which task was not recorded there is no measure of how much humans disagree with each other.
Agreement with that rating bounds the bias only as far as a few informed readers' judgement does, and neither they nor any small panel of graduates is a sample of the world's workers.
The agentic panel does not fill that gap: Claude Opus 5.5 with web search brings evidence about how tasks are done, but it comes from the same developer as the two raters and may share their view of which work suits a model, so its agreement with them is a check of consistency, not of validity.
Three independent human raters on the full 1,500 tasks remain the registered standard and are left for a second version.
The second is circularity: if the same family of models both rates delegability and generates the usage data, agreement between the two could reflect a shared view of the world rather than validation.
We address it by testing convergence against indices built by other methods and other raters \citep{eloundou2024gpts, felten2021occupational, gmyrek2025generative}, by comparing with the Agentic Adoption Index, which uses no model rating \citep{lee2026delegates}, and by reporting the Colombian wage dynamics, which involve no model at all.
The third is instability: the command-line interface exposes no temperature setting, so the same prompt can return different levels on different days.
The test--retest statistics in Section~\ref{sec:metodo} quantify this, and the consensus rule across models dampens it; the raw outputs of every pass are released so that others can recompute.
A fourth risk is framing.
In the pilot, a model told that it is an agent rated tasks as more delegable: the interface's default wrapper raised the mean level from \cifra{diseno.piloto.neutro.nivel} to \cifra{diseno.piloto.envoltorio.nivel} on the same tasks.
The ratings therefore depend on how the rater is addressed, not only on the task.
We fix the neutral configuration for the main pass and report, as a preregistered secondary result, the same validation tasks rated under the default wrapper (Section~\ref{sec:validacion}), so that readers can see how much of the index moves with the frame.
Model-based rating of social facts has a short literature \citep{horton2023large, argyle2023out, korinek2023language}, including evidence that model-simulated survey respondents miss the results of human polls \citep{pew2026silicon}, and the known tendency of models to favour their own outputs \citep{zheng2023judging, panickssery2024llm} is a reason to read every number here as a model's judgement, not a measurement of the world.

\subsection{Preregistration}

The hypotheses, rubric, rater set, call configuration, sample seeds and analysis plan were registered on the Open Science Framework on 21 October 2026, before the main pass, under embargo until 31 January 2027 \citep{espinal2026prereg}.
The registration constrains what we can report as confirmatory; everything else is labelled exploratory.
Deviations from the registered plan, such as the source of the employment weights for unit groups within a reported group, are dated in \texttt{docs/DESVIACIONES.md}.
A preregistration does not make a measurement right; it makes it checkable.

\subsection{Ethics}

An index of delegability describes work in the voice of the machine: which parts of a job it could take, and which it could not.
That framing has costs.
It can reduce a worker to the residue of tasks an agent cannot do, and it can read a barrier such as trust or care as a temporary obstacle rather than as the point of the job.
Whether agents complement or replace the people who do these tasks is a design choice as much as a technical fact \citep{autor2024applying}, and a world with less work is not self-evidently a worse one \citep{danaher2019automation}.
We have tried to keep the four barriers visible rather than fold them into a single score, so that a reader can see why a task stays with a human and not only that it does.
The divide we name is also not symmetrical in its meaning.
In high-income countries, high delegability describes a labour market about to reorganise; in low-income countries, low delegability describes a labour market whose work agents cannot reach, which is not the same as protection.
Who is delegated, who delegates, and who is left outside the scope of delegation are distributional questions, and the atlas is meant to put them on the table, not to settle them.
We use no personal data; all inputs are public statistics, public classifications or anonymised household surveys used under their licences.
The human rating in this version was led by a coauthor as part of his work on the paper, with the help of student raters; no external professional raters were recruited, the student raters are not identified, and no personal data about raters are collected or released.

\section{Conclusion}
\label{sec:conclusion}

\cifra{titular.global.l2}\% of task time in world employment is at least partly delegable to AI agents, and only \cifra{titular.global.l3}\% can be handed over in full.
Agents, as the models judge them, take over parts of tasks far more often than whole tasks, and whole tasks mostly when they take a human about twelve minutes.

The share varies with income, from \cifra{titular.ingreso.bajo.l2}\% in low-income to \cifra{titular.ingreso.alto.l2}\% in high-income countries.
This delegation divide is a matter of composition: poorer countries employ more people in work with a physical barrier, and agents have no body.
It is not evidence that the same task is less delegable in a poorer country, and it is not protection; it describes where agents can and cannot reach.

Two preregistered expectations failed, and both failures are informative.
The adoption gap did not appear: observed AI usage tracks delegability closely across countries ($r$ = \cifra{h2.r.adi_uso}), and middle-income countries do not lag behind their delegable task base.
And at the occupation level ADI correlates with exposure indices above the bound we set in advance (\cifra{h3.r.eloundou} with Eloundou et al., \cifra{h3.r.felten} with Felten et al.).
The occupations an agent could take over are, for the most part, the ones a language model could already speed up.
ADI's contribution is not a new ranking of occupations.
It is delegation levels and barriers on top of exposure, for the world: a scale that separates partial from full delegation, a record of why each task stops short, and a task-level index in ISCO-08 for \cifra{cobertura.paises} countries, which correlates far less with observed usage than with exposure.

The blind human rating agrees with the model consensus at $\kappa$ = \cifra{agree.kappa.humano.modelo} and is the more generous of the two; the agentic panel with web search agrees with the consensus at $\kappa$ = \cifra{agree.kappa.panel.modelo}; and the interface's default wrapper raises Sonnet's mean level from \cifra{envoltorio.sonnet.nivel.neutro} to \cifra{envoltorio.sonnet.nivel.cc}, so the frame in which a model is addressed is part of the measurement.
ADI measures technical delegability with the technology of October 2026, not deployment, and it will move as agents improve; the rubric, the code and the ratings are released so that the index can be rerun.
We release the index for every ISCO-08 unit group, the country atlas, the rubric, the ratings and an MCP server at \url{https://github.com/Cespial/delegation-divide}.

\bibliographystyle{plainnat}\bibliography{referencias}
\appendix
\section{Scoring rubric}
\label{app:rubrica}

The rubric below is the file \texttt{scoring/prompts/rubrica\_v1.md} as frozen at preregistration, reproduced character for character.
Its SHA-256 hash is
\begin{center}\small\texttt{bcd076fb705f0f363f4a7eb2c2a403f16653f22f898eeb7d83bada74a0ea0e46}\end{center}
and the scoring script (\texttt{scoring/puntuar.py}) refuses to run the main pass unless the file on disk matches this value.
The prompt sent to the model is the text from ``You are rating'' onward, with \texttt{\string{occupation\_title\string}} and \texttt{\string{task\_text\string}} replaced by the O*NET occupation title and the task statement; the doubled braces in the output block are Python format escapes and reach the model as single braces.
It is sent as the user message of a call that carries one task, to Claude Sonnet 5.5 and Claude Haiku 5.5, with the fixed one-line system prompt quoted in Section~\ref{sec:evaluadores} and no tools.
The same text, with no model output, is the second sheet of each human rater's workbook; the rater guide is \texttt{validacion/GUIA\_EVALUADORES.md}.

{\tiny
\begin{verbatim}
# ADI scoring rubric v1 (frozen at preregistration)

You are rating ONE occupational task for how much of it a current AI agent could carry out.

**Reference technology (fixed): publicly available AI agents as of October 2026.** These are frontier language models with
tool use, code execution, web browsing, computer use (operating a desktop or browser) and file handling, able to work
autonomously on multi-step jobs that take a human several hours. They have **no physical body or robotics**. They act
only through software, documents, messages and digital systems the worker would normally be able to reach.

OCCUPATION: {occupation_title}
TASK: {task_text}

## Definitions (use exactly these)

- **L0 — No meaningful role.** The agent cannot perform or materially speed up any substantial part of the task.
- **L1 — Augmentation.** A human performs the task; the AI assists step by step (drafts, searches, suggestions). The human
  stays in the loop at each step and produces the output.
- **L2 — Partial delegation.** An agent can execute substantial sub-steps of the task end to end from a goal
  specification. A human still has to perform at least one essential step or review and correct the work before use.
- **L3 — Full delegation.** An agent can complete the whole task end to end, to the quality a competent worker would
  accept, given only the goal and access. The human's role is limited to assigning it and accepting the result.

Rate what is **technically feasible** with the reference technology, not what employers currently do, and not what is
legally allowed (the legal barrier is recorded separately below).

## Output

Return ONLY this JSON (no markdown, no commentary):
{{"level": <0|1|2|3>,
 "share_delegable": <int 0-100, percent of the task's working time an agent could take over>,
 "minutes_human": <int, typical minutes a competent human needs for one instance of the task>,
 "barrier_physical": <true|false, task needs physical presence or manipulation>,
 "barrier_accountability": <true|false, law, licence or professional liability requires a human to perform or sign>,
 "barrier_trust": <true|false, task depends on in-person trust, care or persuasion>,
 "barrier_access": <true|false, task depends on systems, data or places an agent would normally not reach>,
 "confidence": "<LOW|MEDIUM|HIGH>"}}
\end{verbatim}
}

\section{Crosswalk O*NET to ISCO-08}
\label{app:crosswalk}

This appendix documents the bridge from O*NET 30.2 task statements to ISCO-08 unit groups implemented in \texttt{datos/construir\_crosswalk.py}.
Every count below is printed by the script's log and is reproducible from the input files listed in \texttt{datos/MANIFIESTO.jsonl}.

\subsection{Inputs}

\begin{itemize}
  \item O*NET 30.2 \emph{Task Statements} (\cifra{diseno.onet.tareas} statements for \cifra{diseno.onet.ocupaciones} O*NET-SOC 2019 occupations), \emph{Task Ratings} (scales IM, importance, and FT, frequency) and \emph{Occupation Data} (titles) \citep{onet2026database}.
  \item O*NET-SOC 2010 to O*NET-SOC 2019 crosswalk, O*NET Resource Center (CSV; \cifra{diseno.onet10_19.codigos2010} codes of 2010 to \cifra{diseno.onet10_19.codigos2019} codes of 2019).
  \item SOC 2010 to ISCO-08 crosswalk of \citet{hardy2018educational}, Institute for Structural Research (IBS), file \texttt{soc10\_isco08.dta}: \cifra{diseno.ibs.filas} rows linking \cifra{diseno.ibs.soc} six-digit SOC 2010 codes to all \cifra{diseno.ibs.isco} ISCO-08 unit groups.
\end{itemize}

\subsection{Steps}

\begin{enumerate}
  \item \textbf{Task weights within O*NET occupation.} For each task, importance is the mean of the IM scale (1--5) and frequency is the mean category of the FT distribution (1--7, from yearly or less to hourly or more), computed as $\sum_k p_k k / 100$ over the seven category percentages. Missing values are imputed with the median of each scale. The task weight is importance $\times$ frequency, normalised to sum to one within the O*NET-SOC 2019 occupation, following \citet{acemoglu2011skills}.
  \item \textbf{O*NET-SOC 2019 to O*NET-SOC 2010.} The official crosswalk is used in reverse. The six-digit SOC 2010 code is the 2010 O*NET-SOC code with its two-digit suffix removed. An O*NET-SOC 2019 code that descends from several 2010 codes (\cifra{diseno.onet10_19.multiples} cases) divides its mapping weight equally among them.
  \item \textbf{Rescue of SOC 2010 codes absent from the IBS file.} Two codes used by O*NET-SOC 2010 are not in the IBS crosswalk: 19-1020 (biological scientists, broad group used as a detailed code) and 29-2055 (surgical technologists). The rule tries, in order, the ``all other'' residual of the same broad group (code ending in 9), then the other detailed codes of the same broad group in ascending order, then the codes of the same minor group. It assigns 19-1020 to 19-1029, which the IBS file maps to ISCO 2131 (biologists, botanists, zoologists and related professionals), and 29-2055 to 29-2051, mapped to ISCO 3259 (health associate professionals not elsewhere classified). Both are documented here because the second is a group-level approximation rather than a conceptual match.
  \item \textbf{New occupations without a 2010 ancestor.} Four O*NET-SOC 2019 codes were created after the 2010 taxonomy and have no row in the crosswalk: 15-1299.04 (penetration testers), 15-1299.05 (information security engineers), 15-1299.06 (digital forensics analysts) and 15-1299.07 (blockchain engineers). They are assigned by hand to SOC 2010 15-1199 (computer occupations, all other), which the IBS file maps to ISCO 2519 (software and applications developers and analysts not elsewhere classified) and 2529 (database and network professionals not elsewhere classified), with weight one half each.
  \item \textbf{SOC 2010 to ISCO-08.} Each SOC 2010 code maps to $k$ ISCO-08 unit groups with weight $1/k$. In the IBS file \cifra{diseno.ibs.k1} codes map to one group, \cifra{diseno.ibs.k2} to two, \cifra{diseno.ibs.k3} to three and \cifra{diseno.ibs.k4mas} to four or more (maximum \cifra{diseno.ibs.kmax}). The mapping weight is multiplied by the within-occupation task weight to give the weight of each (ISCO group, task) pair, then renormalised within the ISCO group.
  \item \textbf{Aggregation to pairs.} The same task reached through several O*NET occupations is summed into one pair per ISCO group. The result is \cifra{diseno.puente.pares} pairs covering \cifra{diseno.puente.tareas} distinct tasks and \cifra{diseno.puente.isco} unit groups, with a median of \cifra{diseno.puente.tareas_grupo.mediana} tasks per group (minimum \cifra{diseno.puente.tareas_grupo.min}, maximum \cifra{diseno.puente.tareas_grupo.max}), and \cifra{diseno.puente.conjuntos} distinct task sets, since some groups share exactly the same O*NET sources.
\end{enumerate}

\subsection{Exclusions and imputations}

\textbf{Military.} The three armed-forces groups, 0110 (commissioned armed forces officers), 0210 (non-commissioned armed forces officers) and 0310 (armed forces occupations, other ranks), have no O*NET tasks and are excluded from the index and from the country weights.

\textbf{Seven imputed groups.} Seven unit groups receive no task through the chain because their SOC counterparts are residual ``all other'' codes that the IBS file routes elsewhere. Each inherits the task set of its three-digit minor group (all non-imputed unit groups with the same first three digits), with weights renormalised, and is flagged \texttt{imputado = True} so that it can be dropped in the robustness check without imputed groups:

\begin{center}
  \small
  \begin{tabular}{@{}ll@{}}
    \toprule
    ISCO-08 & Unit group \\
    \midrule
    3413 & Religious associate professionals \\
    3511 & Information and communications technology operations technicians \\
    4213 & Pawnbrokers and money-lenders \\
    7133 & Building structure cleaners \\
    8155 & Fur and leather preparing machine operators \\
    8159 & Textile, fur and leather products machine operators not elsewhere classified \\
    9613 & Sweepers and related labourers \\
    \bottomrule
  \end{tabular}
\end{center}

\subsection{Why not the earlier matrix}

The first author's earlier work on augmented human capital \citep{espinal2026augmented, espinal2026measuring} bridged SOC to the Colombian adaptation of ISCO-08 at one digit and distributed tasks downward, so that \cifra{diseno.previo.ocupaciones} four-digit occupations shared \cifra{diseno.previo.conjuntos} task sets and the median task belonged to \cifra{diseno.previo.ocupaciones_por_tarea.mediana} occupations.
That matrix cannot separate a surgeon from a statistician.
The chain above keeps \cifra{diseno.puente.conjuntos} distinct task sets across \cifra{diseno.puente.isco} groups and is the only bridge used in this paper.
The occupation-level comparison indices of Section~\ref{sec:metodo} are carried to ISCO-08 with the same chain, applied at the occupation rather than the task level.

\section{How to cite and use ADI}
\label{app:citar}

\subsection{Citation}

Please cite the paper for any use of the index, the atlas, the rubric or the tools.
\begin{verbatim}
@article{espinal2026delegation,
  title   = {The Delegation Divide: An Open Index of AI-Agent Delegability
             across Occupations and Countries},
  author  = {Espinal Maya, Cristian and Jim{\'e}nez Londo{\~n}o, Santiago},
  journal = {arXiv preprint},
  year    = {2026},
  url     = {https://github.com/Cespial/delegation-divide}
}
\end{verbatim}
The repository carries the same entry in \texttt{CITATION.cff} and at the top of \texttt{README.md}; the dataset and code of this version have their own DOI on Zenodo, \href{https://doi.org/10.5281/zenodo.23552918}{\texttt{10.5281/zenodo.23552918}} (all versions: \texttt{10.5281/zenodo.23552917}), which should be cited together with the paper when the data are used directly.
The index is versioned: this paper describes \texttt{v2026.1}.
Later versions re-rate tasks against a later reference technology and are not interchangeable with this one.

\subsection{Files}

The release contains, under \texttt{salidas/}:
\begin{itemize}
  \item \texttt{adi\_isco.csv}: one row per ISCO-08 unit group with $\mathrm{ADI}_o$, $\mathrm{ADI3}_o$, $\mathrm{ADIs}_o$, the bootstrap interval, the four barrier shares and the number of tasks.
  \item \texttt{adi\_pais.csv}: one row per country with $\mathrm{ADI}_c$ and its variants, the ISCO level used, the reference year, the income group and total employment.
  \item \texttt{adi\_tarea.csv}, with the task statement, the consensus level and each model's level, and the raw model outputs (\texttt{ratings\_sonnet.jsonl.gz}, \texttt{ratings\_haiku.jsonl.gz}), so that any aggregation rule can be recomputed.
  \item \texttt{claims.jsonl}: every number in this paper with its identifier, value, interval and the script that produces it. The LaTeX source reads numbers from this file only.
\end{itemize}

\subsection{Static JSON}

Each occupation and each country is also served as a static JSON document, so that a tool or an agent can fetch one record without downloading the dataset:
\begin{verbatim}
https://cespial.github.io/delegation-divide/adi/v2026.1/isco/2512.json
https://cespial.github.io/delegation-divide/adi/v2026.1/country/COL.json
\end{verbatim}
The documents carry the version, the licence and the citation string in their body.

\subsection{Python package}

\begin{verbatim}
pip install adi-index

import adi
adi.lookup("2512")             # ISCO-08 code
adi.search("software")         # title substring
adi.country("COL")             # ISO 3166-1 alpha-3
adi.top(10)                    # most delegable occupations
adi.cite()                     # BibTeX string
\end{verbatim}
The package ships the index as data and needs no network access; it exposes the same operations as the server below.

\subsection{MCP server}

The Model Context Protocol (MCP) is an open standard that lets an AI agent call external tools \citep{anthropic2024mcp}.
The server \texttt{adi-mcp} (PyPI; listed in the MCP registry) offers four tools:
\begin{description}
  \item[\texttt{lookup\_occupation}] takes an ISCO-08 code, an ESCO occupation identifier or a free-text title and returns the occupation's levels, barriers, intervals and version.
  \item[\texttt{country\_profile}] takes an ISO 3166-1 alpha-3 code and returns the country's index, its variants, the ISCO level and year of the employment data, and the income group.
  \item[\texttt{top\_occupations}] returns the $n$ most delegable occupations at level L2+ or L3.
  \item[\texttt{cite}] returns the BibTeX entry above and the dataset DOI, so that an agent that uses the index can cite it in its output.
\end{description}
A typical configuration for a Claude Code or similar client is:
\begin{verbatim}
{"mcpServers": {"adi": {"command": "uvx", "args": ["adi-mcp"]}}}
\end{verbatim}
Every response includes the citation string.

\subsection{llms.txt}

The repository root and the dataset site carry \texttt{llms.txt} and \texttt{llms-full.txt}, which summarise the paper, the headline number, the file layout and the citation in a form meant to be read by language models.
Agents that index repositories should start there.

\subsection{Licences}

Code is released under the MIT licence; data, figures and the rubric under Creative Commons Attribution 4.0 (CC BY 4.0).
O*NET, ESCO, ILOSTAT and World Bank inputs keep their own licences; O*NET task statements are redistributed verbatim under its CC BY 4.0 licence, and the other inputs only as derived aggregates.

\end{document}